\documentclass[10pt,journal,compsoc]{IEEEtran}
\ifCLASSOPTIONcompsoc
  \usepackage[nocompress]{cite}
\else
  \usepackage{cite}
\fi
\ifCLASSINFOpdf
\else
\fi

\usepackage{subfig}
\usepackage{graphicx} 
\usepackage{booktabs}
\usepackage{multirow}
\usepackage{subfig}
\usepackage{enumitem}
\usepackage{amsmath,amsfonts,amsthm}
\usepackage{bm}
\usepackage[linesnumbered,ruled,vlined]{algorithm2e}
\usepackage{color}
\usepackage{ulem}

\begin{document}
%
\title{Dynamic Graph Neural Networks for\\ Sequential Recommendation}
%
%
%
%

\author{Mengqi~Zhang,~
        Shu~Wu,~\IEEEmembership{Member,~IEEE,}~
        Xueli~Yu,~
        Qiang~Liu,~\IEEEmembership{Member,~IEEE,}~
        Liang~Wang,~\IEEEmembership{~Fellow,~IEEE}
\IEEEcompsocitemizethanks{\IEEEcompsocthanksitem Mengqi Zhang, Shu Wu, Qiang Liu, and Liang Wang are with the Center for Research on Intelligent Perception and Computing (CRIPAC), Institute of Automation, Chinese Academy of Sciences, Beijing 100190, China, and also with the School of Artificial Intelligence, University of Chinese Academy of Sciences, Beijing 101408, China (E-mail: mengqi.zhang@cripac.ia.ac.cn, shu.wu@nlpr.ia.ac.cn, qiang.liu@nlpr.ia.ac.cn, wangliang@nlpr.ia.ac.cn). \protect\\
\IEEEcompsocthanksitem Xueli Yu is with Beijing Institute for General Artificial Intelligence (BIGAI), Beijing, China (E-mail: yuxueli@bigai.ai). 
}
\thanks{(Corresponding author: Mengqi Zhang)}}

%
%

\markboth{Journal of \LaTeX\ Class Files,~Vol.~14, No.~8, August~2015}%
{Shell \MakeLowercase{\textit{et al.}}: Bare Demo of IEEEtran.cls for Computer Society Journals}
%



\IEEEtitleabstractindextext{%
\begin{abstract}
Modeling user preference from his historical sequences is one of the core problems of sequential recommendation. Existing methods in this field are widely distributed from conventional methods to deep learning methods. However, most of them only model users' interests within their own sequences and ignore the dynamic collaborative signals among different user sequences, making it insufficient to explore users' preferences. We take inspiration from dynamic graph neural networks to cope with this challenge, modeling the user sequence and dynamic collaborative signals into one framework. We propose a new method named \emph{Dynamic Graph Neural Network for Sequential Recommendation} (DGSR), which connects different user sequences through a dynamic graph structure, exploring the interactive behavior of users and items with time and order information. Furthermore, we design a Dynamic Graph Recommendation Network to extract user's preferences from the dynamic graph. Consequently, the next-item prediction task in sequential recommendation is converted into a link prediction between the user node and the item node in a dynamic graph. Extensive experiments on three public benchmarks show that DGSR outperforms several state-of-the-art methods. Further studies demonstrate the rationality and effectiveness of modeling user sequences through a dynamic graph. 
\end{abstract}

}

\maketitle

\IEEEdisplaynontitleabstractindextext

%
\IEEEpeerreviewmaketitle

\IEEEraisesectionheading{\section{Introduction}\label{sec:introduction}}

%
%
%
%
\IEEEPARstart{W}{ith} dramatic growth of the amount of information on the Internet, recommender systems have been applied to help users alleviate the problem of information overload in online services, such as e-commerce, search engines, and social media. Recently, several collaborative filtering methods have been proposed, which focus on static user-item interactions \cite{sarwar2001item,he2017neural,wang2019neural}, but ignoring the rich historical sequential information of users. However, user preferences change dynamically over time, varying with the historical interacted items. Therefore, sequential recommendation has attracted lots of attention, which seeks to utilize the sequential information from each user's interaction history to make accurate predictions.

A series of methods have been proposed in the field of sequential recommendation.
For example, the Markov-chain model \cite{rendle2010factorizing} makes recommendation based on $k$ previous interactions. Several RNN-based models \cite{hidasi2015session,quadrana2017personalizing,hidasi2018recurrent} utilize Long Short-term Memory (LSTM) \cite{sak2014long} or Gated Recurrent Unit (GRU) \cite{chung2014empirical} networks to capture sequential dependencies in user sequences. Furthermore, Convolutional Neural Networks (CNN) and Attention Networks are also effective in modeling user sequences. For example, Caser \cite{tang2018personalized} employs convolutional filters to incorporate the order of user interaction. SASRec \cite{kang2018self} and STAMP \cite{liu2018stamp} apply the attention mechanism to model the relationship between items to capture user intent. Recently, Graph Neural Networks (GNNs) \cite{DBLP:conf/iclr/KipfW17,DBLP:conf/iclr/VelickovicCCRLB18} have gained increasing attention. Inspired by the success of GNN in a wide variety of tasks, some GNN-based sequence models \cite{wu2019session,zhang2020personalized,qiu2019rethinking} are proposed, which use improved GNN to investigate the complex item transition relationships in each sequence. 

\begin{figure}[t]
	\centering
	\subfloat[The left figure presents differnt sequences of $u_1$, $u_2$ and $u_3$. Our goal is to predict the next interaction of $u_1$. The figure on the right illustrates the training and testing paradigm of most sequential models.  ]{\includegraphics[scale=0.5]{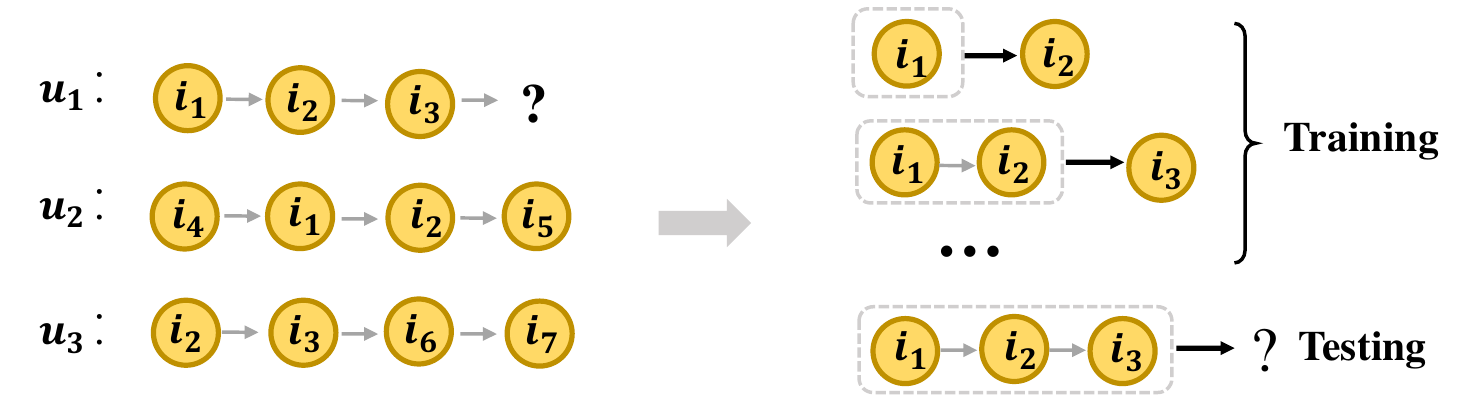}\hspace{5mm}
		\label{fig_first}}\\
	\subfloat[Interaction of user-item graph composed of $u_1$, $u_2$ and $u_3$ at different times. Each edge represents the interaction between user and item, and has time attribute. The node $u_1$ is the target user to predict. The solid line indicates the interaction that has occurred at the current time. The dotted arrow represents the next interactions of $u_1$. The timestamp sequence of $u_1$ is $(t_1, t_2, t_3)$. ]{\includegraphics[scale=0.5]{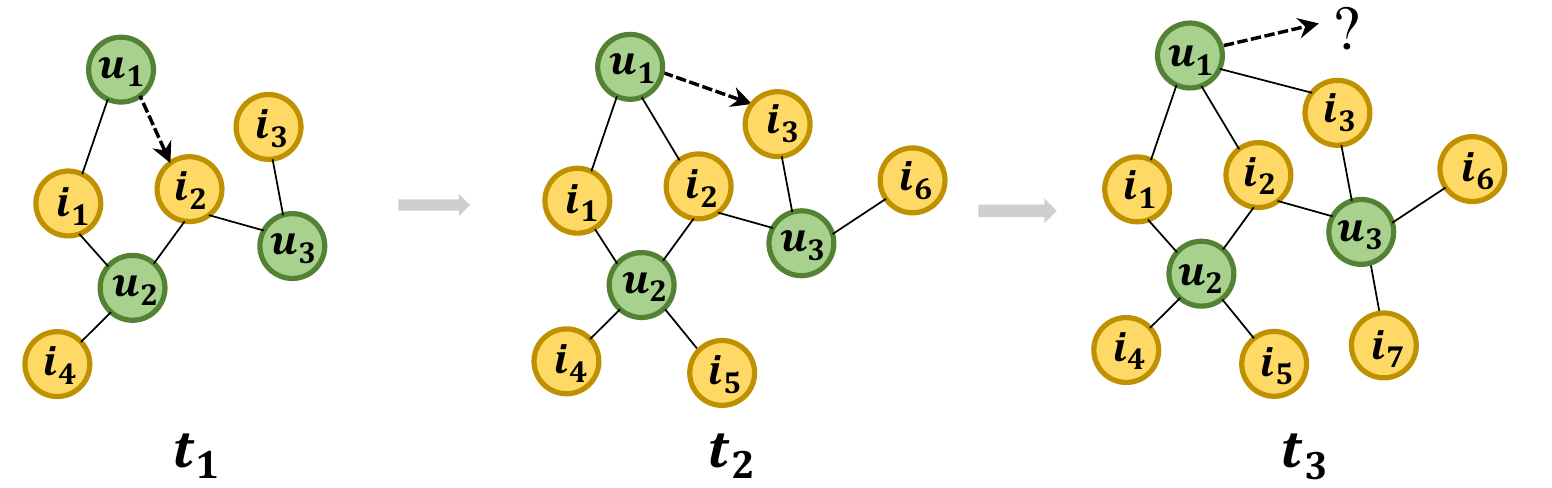}
		\label{fig_thrid}}
	\caption{Illustration of user-item sequential interaction. Figure (a) illustrates the interaction sequences of $u_1$, $u_2$ and $u_3$. Figure (b) is the user-item interaction graph composed of $u_1$, $u_2$ and $u_3$ at different times, which can be seen as a refined representation of Figure (a).}
	\label{ab}
\end{figure}

Although these methods have achieved compelling results, we argue that these methods lack explicit modeling of the \emph{dynamic collaborative signals} among different user sequences, which is mainly manifested in two aspects:

(1) These models do not explicitly leverage the collaborative information among different user sequences, in other words, most of them focus on encoding each user's own sequence, while ignoring the high-order connectivity between different user sequences, as can be seen in Figure \ref{fig_first}, in which the encoding of user sequence during training and testing are all within a single sequence. However, as shown in Figure \ref{fig_thrid}, at $t_3$ time, $u_1$ interact with $i_1$, $i_2$ and $i_3$ directly, and also have high-order connections with $u_2$ and $u_3$ as well as their interactive items. Obviously, the interaction information of $u_2$ and $u_3$ can assist in the prediction of $u_1$'s sequence. This information is overlooked by most of the existing models. 

(2) These models ignore the dynamic influence of the high-order collaboration information at different times. From Figure \ref{fig_thrid}, we can see that the graph formed by $u_1$'s sequence and its high-order associated users and items vary with $t_1$, $t_2$ and $t_3$ time. In this case, the change of $u_1$'s interest is influenced not only by the change of first-order interaction items $i_1$, $i_2$ and $i_3$, but also by the varied high-order connected users and items. Similarly, the semantic information of items may also shift with the change of first- and higher-order relevance.

Consequently, the above two aspects result in the difficulty of accurately capturing user preference in sequential recommendation. To deal with these challenges, two important problems need to be solved: 

(1) \emph{How to dynamically represent user-item interactions with a graph.} The order of interaction between users and items is vital for sequential recommendation. Most exiting methods \cite{wang2019neural} represent user-item interactions as a static bipartite graph and fail to record the interaction order of the user-item pair. So, we need to consider incorporating sequence information or interaction order into the graph flexibly and efficiently.


(2) \emph{How to explicitly encode the dynamic collaborative signal for each user sequence.} For each user sequence, its dynamic associated items and users form a graph structure, which includes more time/order information than conventional static graph. It is not trivial to encode the preference of each user from this dynamic graph.

To this end, inspired by dynamic graph representation learning \cite{trivedi2018dyrep}, we propose a novel method named \emph{Dynamic Graph Neural Network for Sequential Recommendation} (DGSR), which explores interactive behaviors between users and items through dynamic graph. The framework of DGSR is as follows: \emph{firstly}, we convert all user sequences into a dynamic graph annotated with time and order information on edges (Section \ref{section:dynamic graph Construction}). Consequently, the user sequences having common items are associated with each other via \emph{user $\to$ item} and \emph{item $\to$ user} connections. \emph{Secondly}, we devise a sub-graph sampling strategy (Section \ref{section:sub-graph sampling}) to dynamically extract sub-graphs containing user's sequence and associated sequences. \emph{Thirdly}, to encode user's preference from the sub-graph, we design a Dynamic Graph Recommendation Network (DGRN) (Section \ref{graph neural network}), in which a dynamic attention module is constructed to capture the long-term preference of users and long-term character of items, and a recurrent neural module or attention module is further utilized to learn short-term preference and character of users and items, respectively. By stacking multiple DGRN layers, the rich dynamic high-order connectivity information of each user and each item node can be better utilized. \emph{Finally}, our model converts the next-item prediction task into a link prediction task for user nodes (Section \ref{prediction layer}). Extensive experiments conducted on three public benchmark datasets verify the effectiveness of our DGSR method.

To summarize, our work makes the following main contributions:
\begin{itemize}
	\item We highlight the critical importance of explicitly modeling dynamic collaborative signals among user sequences in the sequential recommendation scenario. 
	\item We propose DGSR, a new sequential recommendation framework based on dynamic graph neural networks. 
	\item We conduct empirical studies on three real-world datasets. Extensive experiments demonstrate the effectiveness of DGSR.
\end{itemize}
\section{Related Work}
\subsection{Sequential Recommendation}
Sequential recommendation is to predict the next item based on users' historical interaction sequences. 
Compared with the static recommender system, it usually generates user's representation based on its sequential interactions for prediction. Pioneering works, such as Markov chain-based methods \cite{rendle2010factorizing,he2016fusing} predict the next item based on k-order interaction. Translation-based approaches \cite{he2017translation} model third-order interactions with a TransRec component.

With the development of deep learning, many related works have been proposed for sequential recommendation task. GRU4Rec \cite{hidasi2015session} is the first one to use Recurrent Neural Network (RNN) to session-based recommendation task. Due to the excellent performance of RNN, it has been widely used for sequential recommendation task \cite{yu2016dynamic,quadrana2017personalizing,hidasi2018recurrent}. What's more, Convolution Neural Network (CNN) is also used in sequential recommendation to investigate the different patterns. The CNN-based model Caser \cite{tang2018personalized} applies convolution filters to incorporate different order of users' interactions. Furthermore, the attention network is also a powerful tool applied in the sequential recommendation. NARM \cite{li2017neural} employs the attention mechanism on RNN to capture users' main purposes. STAMP \cite{liu2018stamp} uses a novel attention memory network to efficiently capture both the users’ general interests and current interests.  
SASRec \cite{kang2018self} applies self-attention mechanisms to sequential recommendation problems to explicitly model the relationship between items. More recently, based on SASRec, TiSASRec \cite{li2020time} is proposed to model the absolute positions of items as well as the time intervals between them in a sequence.

In the last few years, graph neural networks (GNNs) have achieved state-of-the-art performance in processing graph structure data. There are also some studies \cite{wu2019session,zhang2020personalized,qiu2019rethinking,xu2019graph,DBLP:conf/aaai/MaMZSLC20} applying GNNs to sequential recommendation. SR-GNN \cite{wu2019session} firstly utilizes the Gated GNNs to capture the complex item transition relationship in session scenario. Based on this work, A-PGNN \cite{zhang2020personalized} combining personalized GNN and attention mechanism is proposed for session-aware scenarios. MA-GNN \cite{DBLP:conf/aaai/MaMZSLC20} employs a memory augmented graph neural network to capture both the long- and short-term user interests. 

Although these GNN-based models have shown promising direction for sequential recommendation, they only focus on modeling user preferences on intra-sequence and ignore the item relationship across sequences. To this end, some models are proposed. For example, HyperRec \cite{wang2020next} adopts hypergraph to model the high-order correlations connections between items within or across sequences. CSRM \cite{wang2019collaborative} considers neighborhood sessions by calculating that of similarity between with current session. DGRec \cite{song2019session} explicitly associate different user sequences through social relationships, but not all data have social relationship attributes. Furthermore, to effectively learn user and item embeddings, THIGE \cite{ji2020temporal} utilizes the temporal heterogeneous graph for next-item recommendation. However, our model processing sequential recommendation task is distinct from the above-mentioned methods. The detailed comparative analysis with these models will be elaborated in Section \ref{discussion}.
\subsection{Dynamic Graph Neural Networks}
Nowadays, graph neural networks have been employed to address different problems, such as node classification \cite{DBLP:conf/iclr/KipfW17,DBLP:conf/iclr/VelickovicCCRLB18}, graph embedding \cite{tang2015line,perozzi2014deepwalk}, graph classification \cite{DBLP:conf/icml/GaoJ19}, recommendation \cite{wu2019session,zhang2020personalized,qiu2019rethinking,xu2019graph,DBLP:conf/aaai/MaMZSLC20,9338436} and so on. 

However, in many applications, the graph data change over time, such as academic network, social network, and recommender system. As a result, a surge of works considers modeling dynamic graphs. DANE \cite{li2017attributed} leverages matrix perturbation theory to capture the changes of adjacency and attribute matrix in an online manner. DynamicTriad \cite{Zhou2018} imposes the triadic closure process to preserve both structural information and evolution patterns of dynamic network. DynGEM \cite{Goyal2018} uses a dynamically expanding deep Auto-Encoder to capture highly nonlinear first-order and second-order proximities of the graph nodes. CTDNE \cite{Nguyen2018} designs a time-dependent random walk sampling method for learning dynamic network embeddings from continuous-time dynamic networks. HTNE \cite{Zuo2018} integrates the Hawkes process into network embeddings to capture the influence of historical neighbors on the current neighbors for temporal network embedding. Dyrep \cite{DBLP:conf/iclr/TrivediFBZ19} utilizes a deep temporal point process model to encode structural-temporal information over graph into low dimensional representations. JODIE \cite{kumar2019predicting} utilizes two types of RNN to model the evolution of different node representations. MTNE \cite{huang20motif} not only integrates the Hawkess process to stimulate the triad evolution process, but also combines attention to distinguish the importance of different motifs. To inductively infer embeddings for both new and observed nodes as the graph evolves, Xu et.al \cite{xu2020inductive} propose the temporal graph attention mechanism based on the classical Bochner's theorem. There are also some works \cite{sankar2020dysat} crop the dynamic graph into a sequence of graph snapshots.

Although some of the above-mentioned dynamic methods are tested on e-commerce data sets, they are not adapt to sequential recommendation scenarios. As far as we know, there is no study to illustrate the sequential recommendation problem from the perspective of dynamic graphs.
\begin{figure*}[ht]
	\centering
	\includegraphics[scale=0.56]{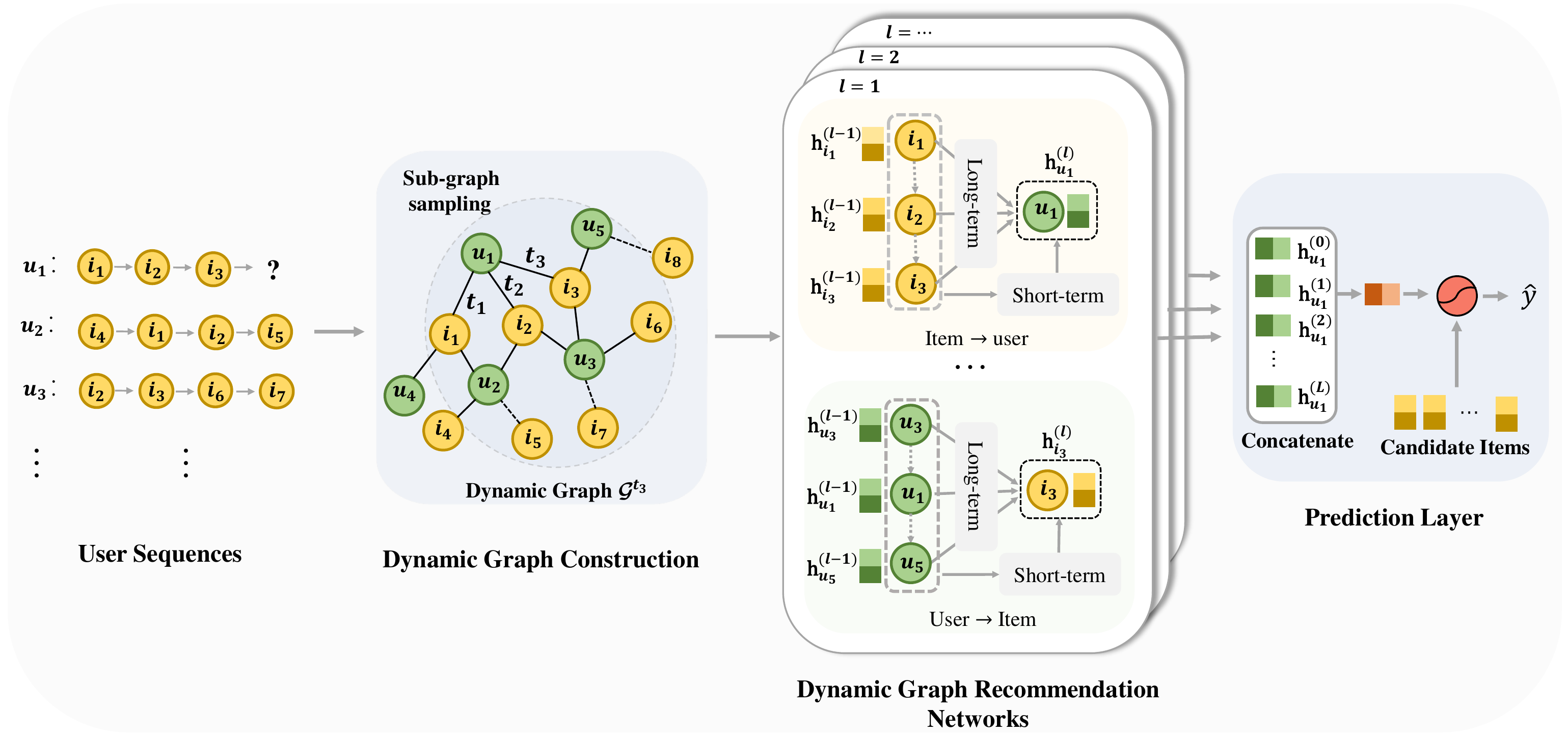}
	\caption{Overview of DGSR framework. Take predicting the next interaction of $u_1$'s sequence $(i_1, i_2,i_3)$ as an example. The corresponding timestamp sequence is $\left(t_1,t_2,t_3\right)$. We first convert $u_1$' sequence and its related sequences into dynamic graph $\mathcal{G}^{t_3}$, each edge represents the interaction between user and item, and has time attribute. The edges represented by the dotted line are interactions that occurred after $t_3$, which is not included in $\mathcal{G}^{t_3}$ (Section \ref{section:dynamic graph Construction}). Then we sample a $m$-order sub-graph $\mathcal{G}^m_{u_1}(t_3)$ from $\mathcal{G}^{t_3}$ (Section \ref{section:sub-graph sampling}). Following this, the well-designed Dynamic Graph Recommendation  Networks propagate and aggregate the information among different user sequences (Section \ref{graph neural network}). Finally, we concatenate user node embedding of each layer for final predication (Section \ref{prediction layer}).}
	\label{fig:frame}
\end{figure*}
\section{Preliminaries}
In this section, we describe problems about sequential recommendation and dynamic graph.
\subsection{Sequential Recommendation}
In the setting of sequential recommendation, let $\mathcal{U}$ and $\mathcal{I}$ represent the set of users and items, respectively. For each user $u \in \mathcal{U}$, its action sequence is denoted as $S^u=\left(i_{1},i_{2},\cdots,i_k\right)$, where $i \in \mathcal{I}$,  $T^u$ $=$ $\left(t_1,t_2,\cdots,t_k\right)$ is the corresponding timestamp sequence of $S^u$. The set of all $S^u$ is denoted as $\mathcal{S}$. The object of sequential recommendation is to predict the next item of ${S^u}$ employing sequence information before time $t_k$ and $t_k$. In general, sequential recommendation task limits the maximum length of $S^u$ to $n$. When $k$ is greater than $n$, taking the most recent $n$ items $\left(i_{k-n},i_{k-n+1},\cdots,i_k\right)$ to make predictions. 

Each user and item can be converted into low-dimensional embedding vector $\mathbf{e}_u$, $\mathbf{e}_i \in \mathbb{R}^d$, respectively, where $u \in \mathcal{U}$ and $i \in \mathcal{I}$, $d$ is the dimension of embedding space. We use the $\mathbf{E}_U \in \mathbb{R}^{|\mathcal{U}|\times d}$ and $\mathbf{E}_I \in \mathbb{R}^{|\mathcal{I}|\times d}$ representing the user embedding and item embedding matrix, respectively. 

\subsection{Dynamic Graph} Generally, there are two types of dynamic graphs \cite{kazemi2020representation}, which are discrete-time dynamic graphs and continuous-time graphs. Our work mainly refers to the continuous-time dynamic graph.

A dynamic network can be defined as $\mathcal{G}=(\mathcal{V}, \mathcal{E}, \mathcal{T})$, where $\mathcal{V}=\{v_1,v_2,\cdots,v_n\}$ is the node set and $e \in \mathcal{E}$ represents the interaction between $v_i$ and $v_j$ at time $t \in \mathcal{T}$, so edge $e_{ij}$ between $v_i$ and $v_j$ is generally represented by triplet $\left(v_i, v_j, t\right)$. In some cases, $t$ can also indicate the order of interactions between two nodes. 
By recording the time or order of each edge, a dynamic graph can capture the evolution of the relationship between nodes. Dynamic graph embedding aims to learn mapping function $f: \mathcal{V} \rightarrow  \mathbb{R}^d$, where $d$ is the number of embedding dimensions. 
\section{Methodology}
We now present the proposed DGSR model, the framework of which is illustrated in Figure \ref{fig:frame}. There are four components in the architecture: 1) \emph{Dynamic Graph Construction} is to convert all sequences of users to a dynamic graph; 2) \emph{Sub-graph Sampling} is to extract sub-graphs which contain user's sequence and its related sequences; 3) \emph{Dynamic Graph Recommendation Networks} (DGRN) contains message propagation mechanism and node update part to encode each user preference from the sub-graph; and 4) \emph{Predication Layer} aggregates the user's refined embeddings learned from DGRN, and predicts which item node will be most likely linked with the user node next. Algorithm \ref{alg:dgsr} provides the pseudo-code of the overall framework.

\subsection{Dynamic Graph Construction}
\label{section:dynamic graph Construction}
In this section, we describe how to convert all user sequences into a dynamic graph. When the user $u$ acts on the item $i$ at time $t$, an edge $e$ is established between $u$ and $i$, and $e$ can be represented by the quintuple $(u,i,t,o_u^i, o_i^u)$. $t$ describes the timestamp when the interaction occurred. Besides, distinguished with the definition of the conventional dynamic graph, $o_u^i$ is the order of $u$$-$$i$ interaction, that is, the position of item $i$ in all items that the $u$ has interacted with. $o_i^u$ refers to the order of $u$ in all user nodes that have interacted with item $i$. For example, $u_1$'s sequence and timestamp sequence are $\left(i_1, i_2, i_3\right)$ and $\left(t_1, t_2, t_3\right)$, respectively.  $u_2$'s sequence and timestamp sequence are $\left(i_2, i_3, i_1\right)$ and $\left(t_4, t_5, t_6\right)$, where $t_1 < t_2 < t_3<t_4<t_5<t_6$. The edges between users and its interaction items can be written as $(u_1, i_1, t_1, 1, 1)$, $(u_1, i_2, t_2, 2, 1)$, $(u_1, i_3, t_3, 3, 1)$, $(u_2, i_2, t_4, 1,2)$, $(u_2, i_1, t_5, 2, 2)$, and $(u_2, i_3, t_6, 3, 2)$. 

Since a large number of user sequences interacted with the same items,
for example, as show in the Figure \ref{fig:frame}, $u_1$ and $u_2$ have common items $i_1$ and $i_2$, and $u_1$ and $u_3$ have common item $i_3$. Consequently, all the quintuples of dataset form a dynamic graph, 
$$\mathcal{G} = \{(u,i,t,o_u^i, o_i^u)| u\in\mathcal{U},i\in\mathcal{V}\}.$$ 
In addition to the interaction time between users and items, $\mathcal{G}$ also records the order information between them. So, our dynamic graph is more suitable for the sequential recommendation task than static graph and conventional dynamic graph. We define our dynamic graph at time $t$ as $\mathcal{G}^{t} \in \mathcal{G}$, which is a dynamic graph composed of all users' interaction sequences at time $t$ and before $t$. For a given user sequence $S^u$ $=$ $\left(i_1,i_2,\cdots,i_k\right)$, where the corresponding timestamp sequence is $T^u$ $=$ $\left(t_1,t_2,\cdots,t_k\right)$, the next item of the predicted sequence $S^u$ is equivalent to predict the item linked to the node $u$ in $\mathcal{G}^{t_{k}}$. 

\subsection{Sub-Graph Sampling}
\label{section:sub-graph sampling}
As the user sequence $S^u$ extending, the number of neighbor sequences of it is growing. Similarly, the scale of the dynamic graph composed of all users is also gradually expanding. It will increase the computational cost and introduce too much noise into the target sequence $S^u$. For efficient training and recommendation, we propose a sampling strategy, which details are shown in Algorithm \ref{alg1}.

Specifically, we first take user node $u$ as the anchor node and select its most recent $n$ first-order neighbors from graph $\mathcal{G}^{t_k}$, that is, the historical items that $u$ has interacted with, written as $\mathcal{N}_u$, where $n$ is the maximum length of user sequence (Line 5, 6, and 8 in Algorithm \ref{alg1}). Next, for each item $i \in \mathcal{N}_u$, we use each of them as an anchor node to sample the set of users who have interacted with them, written as $\mathcal{N}_i$ (Line 11, 12, and 14 in  Algorithm \ref{alg1}). To improve sampling efficiency, we record user and item nodes that have been used to be anchor node to avoid repeated sampling (Line 7 and 13 in Algorithm \ref{alg1}).
Followed by analogy, we can obtain the multi-hop neighbors of node $u$, which could forms $u$'s $m$-order sub-graph $\mathcal{G}_u^m(t_k)$ of $S^u$ ($m$ is hyper-parameter used to control the size of sub-graph). 

After sampling, each sub-graph $\mathcal{G}_u^m(t_k)$ contains the nodes of the sequence $S^u$ and its associated sequences. User and item nodes in these sequences are linked to each other through stacking the \emph{user to item} and \emph{item to user} relationships in $\mathcal{G}_u^m(t_k)$.

\begin{algorithm}
	\caption{Sub-graph Sampling Algorithm} 
	\label{alg1} 
	\SetKwInOut{Input}{Input}\SetKwInOut{Output}{Output}
	\Input{Sequence $S^u$ $=$ $\left(i_1,i_2,\cdots,i_k\right)$, timestamp sequence $T^u$ $=$ $\left(t_1,t_2,\cdots,t_k\right)$, dynamic graph $\mathcal{G}^{t_k}$, and the order of sub-graph $m$.}
	\Output{The $m$-order sub-graph $\mathcal{G}_u^m(t_k)$.}
	\BlankLine
	$//$ Initialization \\
	$\mathcal{U}_m, \mathcal{U}_{temp} \leftarrow \{u\}$, $\mathcal{I}_m, \mathcal{I}_{temp} \leftarrow \{i_1,\cdots, i_k\}$, $j=0$ \\
	$//$ Node sampling \\ 
	\While{j $\leq$ m}{
		\For{$i \in \mathcal{I}_{temp}$}{$\mathcal{U}_{temp}\leftarrow \mathcal{U}_{temp}\mathbf{\cup}\mathcal{N}_i$\\
		}
		$\mathcal{U}_{temp} \leftarrow \mathcal{U}_{temp} \setminus \mathcal{U}_m$\\
		$\mathcal{U}_m \leftarrow \mathcal{U}_m \mathbf{\cup} \mathcal{U}_{temp}$\\
		\If{$\mathcal{U}_{temp}$ = $\emptyset$}{Break}
		\For{$u \in \mathcal{U}_{temp}$}{$\mathcal{I}_{temp}\leftarrow \mathcal{I}_{temp}\mathbf{\cup}\mathcal{N}_u$}
		$\mathcal{I}_{temp} \leftarrow \mathcal{I}_{temp} \setminus \mathcal{I}_m$\\
		$\mathcal{I}_m \leftarrow \mathcal{I}_m \mathbf{\cup} \mathcal{I}_{temp}$\\
		\If{$\mathcal{I}_{temp}$ = $\emptyset$}{Break}
		$j=j+1$
	}
	$//$ Sub-graph generation\\
	$\mathcal{G}_u^m(t_k) = (\mathcal{U}_m, \mathcal{I}_m)$,  $\mathcal{U}_m, \mathcal{I}_m \in \mathcal{G}^{t_k}$
\end{algorithm}


\subsection{Dynamic Graph Recommendation Networks}
\label{graph neural network}
In this section, we design a Dynamic Graph Recommendation Networks (DGRN) to encode each the preference of each user from dynamic contextual information by acting on the sub-graph $\mathcal{G}_u^m(t_k)$. Similar to most GNNs, The DGRN component consists of message propagation and node updating components. For the sake of discussion, we illustrate the message propagation and node updating from $l$$-$$1$-th layer to $l$-th layer of DGRN.

The message propagation mechanism aims to learn the message propagation information from \emph{user to item} and \emph{item to user} in $\mathcal{G}_u^m(t_k)$, respectively. The challenge is how to encode the sequential information of neighbors from user and item perspectives, respectively. The static graph neural networks, such as GCN \cite{DBLP:conf/iclr/KipfW17} and GAT \cite{DBLP:conf/iclr/VelickovicCCRLB18}, are powerful in various graph structure data. However, they fail to capture sequential information of neighbors suitably for each user and item. Some sequence model, such as RNN \cite{hidasi2015session} and Transformer \cite{vaswani2017attention} net, are widely used to model user long- and short-term interest, but they can not deal with graph structured data directly. To this end, we combine the graph neural networks and sequential networks to design a dynamic propagation mechanism. 

\emph{\textbf{From item to user.}} The set of neighbor nodes of the user node $u$ is the items that $u$ has purchased. To update the user node representations in each layer, we need to extract two types of information from the neighbors of each user node, which are \emph{long-term preference} and \emph{short-term preference} respectively. The \emph{long-term preference} \cite{yu2019adaptive} of user reflects his or her inherent characteristics and general preference, which can be induced from the user's all historical items. The \emph{short-term preference} of the user reflects his or her latest interest.


\emph{\textbf{From user to item.}} The set of neighbor nodes of the item node $i$ is the users who purchased it, in which the users are arranged in chronological order. Similar to the user, the neighbors of item also reflect its two types of character. On the one hand, the \emph{long-term character} can reflect the general characters of the item. For example, the \emph{wealthy} people usually buy \emph{high-end} cosmetics. On the other hand, \emph{short-term character} reflects the newest property of item. For example, many non-sports enthusiasts may also buy jerseys or player posters during the World Cup. This part of consumers' behavior means the positioning of soccer equipment has changed from professionalism to universality in this period. However, most existing sequential recommendation methods fail to explicitly capture the impact of user nodes on the item node. To settle this problem, we also consider the message propagation from user to item.

\subsubsection{Message Propagation Mechanism} 
In this subsection, we discuss the message propagation mechanism of DGRN, which includes the encoding of long-term and short-term information.

\noindent \emph{\textbf{Long-term Information.}} To capture the long-term information of each node from its neighbors, we reference graph neural networks and recurrent neural network, which explicitly consider the relationship of nodes with their neighbors and sequence dependence of neighbors, respectively. Furthermore, we also design an order-aware attention mechanism that is more suitable for dynamic sequential recommendation.

\begin{itemize}
	\item \textbf{Graph Convolution Neural Networks}. GCN \cite{DBLP:conf/iclr/KipfW17} is an intuitive approach that aggregate all neighbor node embedding directly:
	\begin{align}
	\mathbf{{h}}_u^{L} &= \frac{1}{|\mathcal{N}_u|}\sum_{i \in{\mathcal{N}_u}}\mathbf{W_1}^{(l-1)}\mathbf{h}_{i}^{(l-1)},\label{1}\\
	\mathbf{{h}}_i^{L} &= \frac{1}{|\mathcal{N}_i|}\sum_{i \in{\mathcal{N}_i}}\mathbf{W_2}^{(l-1)}\mathbf{h}_{u}^{(l-1)},\label{2}
	\end{align}
	where $\mathbf{W_1}^{(l-1)}$, $\mathbf{W_2}^{(l-1)}$ $ \in \mathbb{R}^{d\times d}$  are encoding matrix parameters of item and user in the $l$$-$$1$-th layer, where $|\mathcal{N}_u|$ and $|\mathcal{N}_i|$ is the number of $u$'s and $i$'s neighbor nodes.
\end{itemize}

\begin{itemize}
	\item \textbf{Recurrent Neural Networks}, such as GRU net, is an effective network to model the sequence dependencies. So, we utilize GRU net to calculate the long-term preference/character for user/item nodes from their neighbors, which is computed as
	\begin{align}
	\mathbf{h}^{L}_u &= \operatorname{GRU}_U^{(l)}\left(\mathbf{h}^{(l-1)}_{i_1},\cdots, \mathbf{h}_{i_{|\mathcal{N}_u|}}^{(l-1)}\right), i\in {\mathcal{N}_u},\\
	\mathbf{h}^{L}_i &= \operatorname{GRU}_I^{(l)}\left(\mathbf{h}^{(l-1)}_{u_1}, \cdots, \mathbf{h}_{u_{|\mathcal{N}_i|}}^{(l-1)}\right), u\in {\mathcal{N}_i},
	\label{eq6}
	\end{align}
	where $(\mathbf{h}^{(l-1)}_{i_1},\cdots, \mathbf{h}_{i_{|\mathcal{N}_u|}}^{(l-1)})$ and $(\mathbf{h}^{(l-1)}_{u_1}, \cdots, \mathbf{h}_{u_{|\mathcal{N}_i|}}^{(l-1)})$ are into GRU in chronological order.
\end{itemize}

\begin{itemize}
	\item \textbf{Dynamic Graph Attention Mechanism}. In general, the GNN models focus on explicitly capturing the relationship between the central and neighboring nodes while ignoring the sequence information among neighbors. The sequence model is the opposite. To effectively differentiate the influence of different items and  make full use of the user-item interaction order information, we combine graph attention mechanism and the encoding of sequence information to define a dynamic attention module (DAT). 
	
	
	Specifically, for each interaction quintuple $(u,i, t, o_u^i, o_i^u)$, we define $r_u^i$ as the relative-order of item $i$  to the last item in the neighbors of the user node, i.e. $r_u^i=$ $|\mathcal{N}_u|-o_u^i$. For each discrete values $r$, we assign a unique $\mathbf{p}_{r}^K \in \mathbb{R}^d$ parameter vector as the relative-order embedding to encode the order information. Then, the attention coefficients between $\mathbf{h}_u^{(l-1)}$ and its neighbor node representation $\mathbf{h}_i^{(l-1)}$  are influenced by $\mathbf{p}_{r_i}^K$. So, we define a relative order-aware attention mechanism to differentiate the importance weight $e_{ui}$ of items to the user, taking $l$$-$$1$-th layer node embedding $\mathbf{h}_u^{(l-1)}$ and $\mathbf{h}_i^{(l-1)}$ as the input, formulated as
	\begin{equation}
	e_{ui}=\frac{\left(\mathbf{W_2}^{(l-1)}\mathbf{h}_u^{(l-1)}\right)^\mathrm{T}\left(\mathbf{W_1}^{(l-1)}\mathbf{h}_i^{(l-1)}+\mathbf{p}_{r_u^i}^K\right)}{\sqrt{d}},
	\label{eq1}
	\end{equation}
	where $\mathbf{h}_u^{(0)}$ and $\mathbf{h}_i^{(0)}$ are the user embedding $\mathbf{e}_u$ and the item embedding $\mathbf{e}_i$, respectively. $d$ is the dimension of the embeddings, the scale factor $\sqrt{d}$ is to avoid exceedingly large dot products and speed up convergence. The weighting scores between user and its neighbors are obtained via the softmax function:
	\begin{equation}
	\alpha_{ui} = \operatorname{softmax}({e}_{ui}).
	\label{eq2}
	\end{equation}
	Thus, the \emph{long-term preference} of user can be obtained by aggregating the information from its all neighbors adaptively:
	\begin{equation}
	\mathbf{{h}}_u^{L} = \sum\nolimits_{i \in{\mathcal{N}_u}}\alpha_{ui}\left(\mathbf{W_1}^{(l-1)}\mathbf{h}_{i}^{(l-1)}+\mathbf{p}^V_{r_u^i}\right), 
	\label{eq5}
	\end{equation}
	where $\mathbf{p}_{r_u^i}^V \in \mathbb{R}^d$ is relative-order embedding to capture the order information in user message aggregation.
	
	Similarly, the long-term character of item can be calculated by,
	\begin{align}
	\mathbf{{h}}_i^{L} &= \sum\nolimits_{u \in{\mathcal{N}_i}}\beta_{iu}\left(\mathbf{W_2}^{(l-1)}\mathbf{h}^{(l-1)}_{u} + \mathbf{p}^V_{r_i^u}\right),\label{eq9} 
	\end{align}
	where 
	\begin{align}
	\beta_{iu} &= \operatorname{softmax}(e_{iu}), \\
	e_{iu} &=\frac{\left(\mathbf{W_1}^{(l-1)}\mathbf{h}_i^{(l-1)}\right)^\mathrm{T}\left(\mathbf{W_2}^{(l-1)}\mathbf{h}_u^{(l-1)}+\mathbf{p}_{r_i^u}^K\right)}{\sqrt{d}},
	\end{align}
	and $r_i^u=\lvert \mathcal{N}_i \rvert-o_i^u$, $\mathbf{p}_{r_i^u}^V \in \mathbb{R}^d$ is relative-order embedding to capture the order information in item message aggregation.
\end{itemize}


\noindent \emph{\textbf{Short-term Information.}}
In recommender system, the \emph{short-term information} of the user reflects his or her latest interest, many work \cite{liu2018stamp} utilize the last interaction item embedding as user's short-term embedding, but this ignores the reliance on historical information. To this end, we consider attention mechanism to model the explicit effectiveness between last interaction with historical interactions.


\begin{itemize}
	\item  \textbf{Attention Mechanism}. We consider the attention mechanism between the last item/user with each historical item/user:
	\begin{align}
	\mathbf{h}_u^S &= \sum\nolimits_{i\in \mathcal{N}_u}\hat{\alpha}_{ui}\mathbf{h}_{i}^{(l-1)},\\
	\mathbf{h}_i^S &= \sum\nolimits_{u\in {\mathcal{N}_i}}\hat{\beta}_{iu}\mathbf{h}_{u}^{(l-1)},
	\end{align}
	where attention coefficient $\hat{\alpha}_{i_k}$ and $\hat{\beta}_{u_k}$ can be calculated by, 
	\begin{align}
	\hat{\alpha}_{ui} &=  \operatorname{softmax}\left(\frac{\left(\mathbf{W}^{(l-1)}_3\mathbf{h}_{i_{|\mathcal{N}_u|}}^{(l-1)}\right)^\mathrm{T}\left(\mathbf{W}_2^{(l-1)}\mathbf{h}_{i}^{(l-1)}\right)}{\sqrt{d}}\right),\\
	\hat{\beta}_{iu} &=  \operatorname{softmax}\left(\frac{\left(\mathbf{W}_4^{(l-1)}\mathbf{h}_{u_{|\mathcal{N}_i|}}^{(l-1)}\right)^\mathrm{T}\left(\mathbf{W}_1^{(l-1)}\mathbf{h}_{u}^{(l-1)}\right)}{\sqrt{d}}\right),
	\end{align}
	where parameters $\mathbf{W}_3$ and  $\mathbf{W}_4$ $\in \mathbb{R}^{d}$ is to control the weight of last interaction.
	
\end{itemize}

\subsubsection{Node updating}
In this stage, we aggregate the long-term embedding, short-term embedding, and the previous layer embedding to update the node's representation of $\mathcal{G}_u^m(t_k)$. 

\noindent \emph{\textbf{User node updating.}} For user node, the representation updating rule from $l$$-$$1$-th layer to $l$-th layer can be formulated as 
\begin{equation}
\mathbf{h}_u^{(l)} = \operatorname{tanh}\left(\mathbf{W_3}^{(l)}\left[\mathbf{{h}}_u^{L} \parallel \mathbf{h}^{S}_u \parallel\mathbf{h}_u^{(l-1)}\right]\right),
\label{eq7}
\end{equation}
where $\mathbf{W_3}^{(l)}\in \mathbb{R}^{d\times 3d}$ is a user update matrix to control the information of $\mathbf{{h}}_u^{L}$, $\mathbf{h}^{S}_u$, and $\mathbf{h}_u^{(l-1)}$. 

\noindent \emph{\textbf{Item node updating.}} Analogously, the item representation updating rule is
\begin{align}
&\mathbf{h}^{(l)}_i = \operatorname{tanh}\left(\mathbf{W_4}^{(l)}\left[\mathbf{{h}}_i^{L} \parallel \mathbf{h}^{S}_i \parallel \mathbf{h}^{(l-1)}_i\right]\right),\label{eq8}
\end{align}
where $\mathbf{W_4}^{(l)}\in \mathbb{R}^{d\times 3d}$ is a item update matrix to control the information reservation of $\mathbf{{h}}_i^{L}$, $\mathbf{h}^{S}_i$, and $\mathbf{h}_i^{(l-1)}$.

\subsection{Recommendation and Optimization}
\label{prediction layer}
In our model, predicting the next interaction of $S_u = \left(i_1,i_2,\cdots,i_k\right) $ is equivalent to predicting the link of user node $u$ of sub-graph $\mathcal{G}_u^m(t_k)$. In this subsection, we design the link prediction function to determine the items that the user may interact with next.

After acting  $L$-layers DGRN on $\mathcal{G}_u^m(t_k)$, we obtain the multiple embedding $\{\mathbf{h}_u^{(0)},\mathbf{h}_u^{(1)},\cdots,\mathbf{h}_u^{(L)}\}$ of $u$ node, which includes user embedding  $\mathbf{h}_u^{(l)}$ in each layer. The user embedding in different layers emphasizes the various user preferences \cite{wang2019neural}. As a result, we concatenate user multiple embeddings to get the final embedding for node $u$:
\begin{equation}
\mathbf{h_u} =\mathbf{h}_u^{(0)} \parallel \mathbf{h}_u^{(1)}\cdots\parallel \mathbf{h}_u^{(L)}.
\end{equation}
For a given candidate item $i \in \mathcal{I}$, the link function is defined as
\begin{equation}
\mathbf{s}_{ui}=\mathbf{h_u} ^\mathrm{T}\mathbf{W_P}\mathbf{e}_i,
\end{equation}
where vector $\mathbf{s}_u = (\mathbf{s}_{u1}, \mathbf{s}_{u2},\cdots, \mathbf{s}_{u|\mathcal{I}|})$   represents the score vector of $u$ for each candidate item. $\mathbf{W_P}\in \mathbb{R}^{(L+1)d\times d}$ is the trainable transformation matrix. 
\begin{algorithm}[t]
	\caption{The DGSR framework (forward propagation)} 
	\label{alg:dgsr} 
	\SetKwInOut{Input}{Input}\SetKwInOut{Output}{Output}
	\Input{$S^u$ $=$ $\left(i_1,i_2,\cdots,i_k\right)$, timestamp sequence $T^u$ $=$ $\left(t_1,t_2,\cdots,t_k\right)$, all sequences of users, and DGRN layer number $L$.}
	\Output{The next item $i_{k+1}$ of $S^u$.}
	\BlankLine
	$//$ Dynamic Graph Construction \\
	Convert all user sequences into a dynamic graph $\mathcal{G}$\\
	$//$ {The sub-graph of generation for $S^u$} \\ 
	Run the \textbf{Algorithm \ref{alg1}} to generate $\mathcal{G}_u^m(t_k)$ from $\mathcal{G}^{t_k}$\\
	$//$ {The initialization of node representation } \\ 
	$\mathbf{h}_u^{(0)} \leftarrow \mathbf{e}_u,  \mathbf{h}_i^{(0)} \leftarrow \mathbf{e}_i$, $\forall{u, i}\in \mathcal{G}_u^m(t_k)$  \\
	$//$ {The updte of user and item by DGRN}\\
	\For{$l \in [1:L]$}{
		$\mathbf{h}_u^{(l)}, \mathbf{h}_i^{(l)} \leftarrow \operatorname{DGRN} (\mathbf{h}_u^{(l-1)}, \mathbf{h}_i^{(l-1)}, \mathcal{G}_u^m(t_k)):$\\
		$\mathbf{h}_u^{(L)},\mathbf{h}_i^{(L)}\leftarrow$ Long-term Information Encoding\\
		$\mathbf{h}_u^{(S)},\mathbf{h}_i^{(S)}\leftarrow$ Short-term Information Encoding\\
		$\mathbf{h}_u^{(l)}\leftarrow \operatorname{tanh}\left(\mathbf{W_3}^{(l)}\left[\mathbf{{h}}_u^{L} \parallel \mathbf{h}^{S}_u \parallel\mathbf{h}_u^{(l-1)}\right]\right)$ \\
		$\mathbf{h}^{(l)}_i \leftarrow \operatorname{tanh}\left(\mathbf{W_4}^{(l)}\left[\mathbf{{h}}_i^{L} \parallel \mathbf{h}^{S}_i \parallel \mathbf{h}^{(l-1)}_i\right]\right)$
	}
	$//$ {The prediction of next item.}\\
	$\mathbf{h_u} =\mathbf{h}_u^{(0)} \parallel \mathbf{h}_u^{(1)},\cdots,\parallel \mathbf{h}_u^{(L)}$ \\
	Next item $\leftarrow \operatorname{argmax}\limits_{i\in \mathcal{V} }(\mathbf{h_u} ^\mathrm{T}\mathbf{W_P}\mathbf{e}_i)$
\end{algorithm}

To learn model parameters, we optimize the cross-entropy loss. The normalized vector of user $u$'s score for candidate item is 
\begin{equation}
\mathbf{\hat{y}}_u= \operatorname{softmax}(\mathbf{s}_u).
\end{equation}
The objective function is as follows:
\begin{equation}
Loss=-\sum_{\mathcal{S}}\sum_{i=1}^{\lvert \mathcal{I} \rvert}\mathbf{y}_{ui}\operatorname{log}(\mathbf{\hat{y}}_{ui})+(1-\mathbf{y}_{ui})\operatorname{log}(1-\mathbf{\hat{y}}_{ui})+\lambda \|\Theta\|_2,
\end{equation}
where $\mathbf{y}_u$ denotes the one-hot encoding vector of the ground truth items for the next interaction of $S^u$. $\Theta$ denotes all model parameters, $\| \cdot \|_2$ is $L_2$ norm. $\lambda$ is to control regularization strength.

\subsection{Model Discussions}
\label{discussion}
This subsection compares and analyzes our DGSR with some representative sequential recommendation models.

Some sequence models encoding user preference only depends on its intra-sequence, and does not explicitly utilize other sequence information, such as TiSASRec\cite{li2020time}, SR-GNN \cite{wu2019session}, and HGN \cite{ma2019hierarchical}, which can be viewed as special cases of our DGSR. Specifically, within the one layer DGRN net, we can replace our current setting with some complex network, self-attention net, GGNN net, or Gated net in message propagation mechanism of \emph{item $\to$ user}, and disable the message propagation and node update of \emph{item $\to$ user}. Then, $\mathbf{h}_u^{(1)}$ is treated as $u$'s final preference representation. So, as a new framework, our model can fuse nearly all single-sequence models by modifying the message propagation mechanism part.

There are also some models \cite{ji2020temporal, wang2020next, wang2019collaborative} which are designed to utilize cross sequence information or capture the item relations between different sequences. However, They have many  differences and limitations compared with DGSR. For example, CSRM \cite{wang2019collaborative} considers neighborhood sequences by directly calculating the similarity between them and target sequence but fails to utilize the fine-grained interaction information of users, including the interaction-order between each item with all users that interact with it. Compared with that, our model measures the similarity between different sequences based on the well-designed message passing mechanism, which could improve the utilization of interactions between users and items. HyperRec \cite{wang2020next} adopts hypergraph to associate the high-order correlations connections between items in each period. Nonetheless, hypergraph is a rough way to model user-item interaction, which results in much-refined information being neglected, such as the explicit order information in each cross sequence. The dynamic graph constructed by our DGSR can be more flexible to represent richer interaction information. In social recommendation, DGRec \cite{song2019session} explicitly associate different user sequences through social attribute information, but not all data have social relationship attributes in sequential recommendation scenario. Our model can also explicitly associate different user sequences without relying on other auxiliary information. 



\begin{table}[t]
	\centering
	\caption{The statistics of the datasets.}
	\resizebox{0.9\columnwidth}{!}
	{\begin{tabular}{ccccc}
			\toprule
			\bf Datasets & \bf Beauty & \bf Games & \bf CDs \\
			\midrule
			\bf $\#$ of Users & 52,024     & 31,013 & 17,052 \\
			\bf $\#$ of Items & 57,289    & 23,715 & 35,118 \\
			\bf $\#$ of Interactions & 394,908 & 287,107& 472,265  \\
			\bf Average length & 7.6    &9.3  &27.6  \\   
			\bf Density  &0.01\%  &0.04\%  &0.08\% \\
			\bottomrule
	\end{tabular}}
	\label{tab:dataset}
\end{table}

\begin{table*}[h]
	\centering
	\caption{Performance of DGSR and compared methods in terms of Hit@10 and NDCG@10. The best results is boldfaced. The underlined numbers is the second best results. "Gain" means the improvement over the best compared methods.}
	\resizebox{2\columnwidth}{!}{
		\begin{tabular}{cccccccccccccc}
			\toprule
			\bf Datasets&\bf Metric&  \bf BPR-MF & \bf FPMC & \bf GRU4Rec+ & \bf Caser & \bf SASRec &\bf SR-GNN& \bf HGN & \bf TiSASRec&  \bf{HyperRec} &\bf DGSR &\bf Gain \\ \midrule 
			& {NDCG@10}   & 21.83  & 28.91  &  26.42 & 25.47 & 32.19&32.33 & \underline{32.47}&   30.45 & 23.26& \textbf{35.90} & 10.56\%  \\
			\multirow{-2}{*}{Beauty} & {Hit@10} & 37.75 & 43.10 & 43.98 & 42.64  &  48.54& 48.62 &  \underline{48.63} & 46.87 &34.71&\textbf{52.40}    &  7.75\%    \\ \midrule 
			\multirow{2}{*}{Games}     
			& {NDCG@10}   &   28.75 &  46.80  &  45.64  &   45.93  &    \underline{53.60}&53.25   & 49.34 &   50.19  &48.96&\textbf{55.70} &3.92\%    \\
			& {Hit@10} &      37.75 &   68.02     &  67.15  & 68.83 & \underline{73.98}& 73.49 & 71.42 & 71.85 &71.24&\textbf{75.57} &2.15\% \\ \midrule 
			\multirow{2}{*}{CDs}  
			& {NDCG@10}   &  36.26  & 33.55 &  44.52 & 45.85  & 49.23& 48.95& \underline{49.34} & 48.97 &47.16&\textbf{51.22} &3.81\%  \\
			& {Hit@10} &     56.27  & 51.22 &  67.84 & 68.65  & 71.32& 69.63& \underline{71.42} & 71.00  &71.02&\textbf{72.43} & 1.41\%   \\ \bottomrule
	\end{tabular}}
	\label{table:performance}
\end{table*}

\section{Experiments}
In this section, we perform experiments on three real-world datasets to evaluate the performance of our model. We aim to answer the following questions through experiments. 
\begin{itemize}
	\item \textbf{RQ1}: How does DGSR perform compared with state-of-the-art sequential recommendation methods?
	\item \textbf{RQ2}: How effective is the dynamic graph recommendation networks component in DGSR?
	\item \textbf{RQ3}: What are the effects of different hyper-parameter settings (DGRN layer number, sub-graph sampling size, maximum sequence length, and the embedding size) on DGSR.
\end{itemize}
\subsection{Datasets}
To evaluate the effectiveness of our model, we conduct experiments on three \textbf{Amazon}\footnote{http://jmcauley.ucsd.edu/data/amazon} datasets from real-world platforms \cite{mcauley2015image}: \emph{Amazon-CDs}, \emph{Amazon-Games}, and \emph{Amazon-Beauty}. These datasets are widely used in evaluating sequential recommendation methods and are varying in terms of domains, sizes, and sparsity. 

All of these datasets contain the timestamps or specific dates of interactions. For all datasets, we treat the presence of a review or rating as implicit feedback and discard users and items with fewer than five related actions \cite{kang2018self}. After processed, the data statistics are shown in Table \ref{tab:dataset}. For each user sequence, we use the most recent item for testing, the second recent item for validation, and the remaining items for the training set. To fully capture the dynamic collaborative signals, we segment each sequence $S^u$ into a series of sequences and labels. For example, for an input $S^u=\left(i_{1},i_{2}, i_3, i_4\right)$ and  $T^u=\left(t_{1},t_{2}, t_3, t_4\right)$, we generate sequences and labels as $[i_1]\to i_2$, $[i_1, i_2]\to i_3$ and $[i_1, i_2, i_3]\to i_4$. Then, the corresponding sub-graph and the node to linked are $(\mathcal{G}_u^m(t_1), i2)$,  $(\mathcal{G}_u^m(t_2), i3)$, and $(\mathcal{G}_u^m(t_3), i_4)$. These processing can be take before training and testing.

\begin{table*}[t]
	\centering
	\caption{Performance of  compared with different model variants in terms of NGCD@10 and Hit@10 (``$-$" indicates DGSR does not consider the setting of this part).}
	\resizebox{0.9\textwidth}{!}{
		\begin{tabular}{cccccccccccc}
			\toprule
			\multirow{2.5}[0]{*}{Variants}&\multicolumn{2}{c}{Ablation} & \multicolumn{2}{c}{Beauty}  & \multicolumn{2}{c}{Games} & \multicolumn{2}{c}{CDs} \\
			\cmidrule(lr){2-3} \cmidrule(lr){4-5} \cmidrule(lr){6-7} \cmidrule(lr){8-9}
			&Long-term &Short-term& NDCG@10 & Hit@10 &NDCG@10 & Hit@10 &NDCG@10 & Hit@10 \\
			\midrule 
			DGSR-G&{GCN}& -- & 33.75& 49.94 & 53.44 & 73.23 &48.66&70.43\\
			DGSR-R&{RNN}&--  &34.81 & 50.90& 54.70 & 74.73 &49.57 & 71.22\\
			DGSR-D&{DAT}&--  &35.25 & 51.36 & 55.12 & 74.83 &49.66 &71.26 \\
			\midrule
			DGSR-L& -- & {Last} &  30.87  & 46.13  &  52.43  & 72.18   &46.23 &67.38\\
			DGSR-A& -- & {ATT}  & 34.76   & 51.00  & 54.30   & 74.32   &48.78 &70.09   \\
			\midrule 
			DGSR-GL&{GCN} &Last &  35.24   &   51.18 & 54.76  &   74.58   &49.62 &70.76  \\
			DGSR-RL&{RNN} &Last & 35.47   &   51.68  & 54.86 &   74.84   &  50.26&71.24 \\
			DGSR-DL&{DAT} &Last &  35.62 &    51.92   & 55.53 &  75.07 &    50.72&72.06\\
			\midrule 
			DGSR-GA&{GCN} &ATT &35.00  &  51.05 &54.97 & 74.78  &  50.05 & 71.46   \\
			DGSR-RA&{RNN}&ATT  &35.17  &  51.46 &55.02 & 74.88  & 51.19  & \textbf{72.55}\\
			DGSR-DA&{DAT}&ATT  &\textbf{35.90}  &\textbf{52.40} & \textbf{55.70} & \textbf{75.57} &\textbf{51.22}& 72.43 \\	
			\bottomrule
		\end{tabular}
	}
	\label{ablation study}
\end{table*}

\subsection{Experiment Settings}
\subsubsection{Compared Methods} To demonstrate the effectiveness, we compare our proposed DGSR with the following methods:
\begin{itemize}
	\item \textbf{BPR-MF} \cite{rendle2009bpr}, a matrix factorization based model that learns pairwise personalized ranking from user implicit feedback.
	\item \textbf{FPMC} \cite{rendle2010factorizing}, a model that combines matrix factorization and first-order Markov Chains to capture users' long-term preferences and item-to-item transitions.
	\item \textbf{GRU4Rec+} \cite{hidasi2018recurrent}, an improved RNN-based model that adopts a different loss function and sampling strategy on Top-$K$ recommendation.
	\item \textbf{Caser} \cite{tang2018personalized}, a CNN-based model capturing high-order Markov chains by applying convolution operations on the embedding of the L recent items.
	\item \textbf{SASRec} \cite{kang2018self}, a self-attention-based model to identify relevant items for predicting the next item.
	\item \textbf{SR-GNN} \cite{wu2019session}, a GNN-based model to capture the complex transition relationships of items for the session-based recommendation.
	\item \textbf{HGN} \cite{ma2019hierarchical}, a sequence model that contains feature gating, instance gating, and instance gating modules to select important features and explicitly capture the item relations.
	\item \textbf{TiSASRec} \cite{li2020time}, interval aware self-attention based model, which models both the absolute positions as well as the time intervals between them in a sequence. 
	\item \textbf{HyperRec} \cite{wang2020next}, a hypergraphs based model, which adopts hypergraph to capture multi-order connections between items for next-item recommendation.
\end{itemize}

\subsubsection{Evaluation Metrics} We adopt two widely-used metrics \cite{kang2018self}, Hit@$K$ and NDCG@$K$, to evaluate all methods. Hit@$K$ indicates the proportion of the ground-truth items among the top@$K$ items, while NDCG@$K$ is position-aware metric, and higher NDCG means target items tend to have more top rank positions. Following \cite{li2020time,kang2018self}, for each test sample, we randomly sample 100 negative items, and rank these items with the ground-truth item. We evaluate Hit@$K$ and NDCG@$K$ based on these 101 items. By default, we set $K$=10. 

\subsubsection{Parameter Setup}
We implement our DGSR model in \emph{DGL} Library\footnote{https://www.dgl.ai/}\cite{wang2019dgl}. The embedding size is fixed to 50 for all methods. The maximum sequence length $n$ is set to 50. The optimizer is the Adam optimizer \cite{Kingma2014Adam}, the learning rate is set to 0.01. Batch size is 50. $\lambda$ is 1e-4. We set the order of sub-graph sampling $m$ to 4. The DAN layer number $L$ is set to 3 for Beauty and CDs, 2 for Games. We run the evaluation four times with different random seeds and report the mean value of each method. For the compared methods, we use the default hyperparameters except for dimensions. All experiments are conducted on a computer server with eight NVIDIA GeForce RX2080Ti (11GB) and four Intel Xeon E5-2660 v4 CPUs.

\subsection{Performance Comparison (RQ1)} 
\label{performance}
We first report the performance of all the methods. Table \ref{table:performance} summarizes the performance comparison results. The following observation can be obtained:
\begin{itemize}
	\item DGSR achieves the best performance on three datasets with most evaluation metrics. In particular, DGSR improves over the strongest baselines w.r.t NDCG@10 by $10.56\%$, $3.92\%$, $3.81\%$ in Beauty, Games, and CDs, respectively. Notably, Beauty is the most sparse and short dataset, so many users and items only have a few interactions. In our model, the high-order connectivity of a dynamic graph alleviates this issue. So, there is a significant improvement in Beauty. By stacking the DGRN layers, DGSR can utilize cross-sequences information explicitly to provide more auxiliary information for prediction. While TiSASRec, HGN, SR-GNN, and SASRec only encode each sequence independently as the user's dynamic interest representation. Significantly, HyperRec utilizes many correlated user interaction information, but performs worse than our DGSR, especially on Beauty and Games. We believe that the reason is HyperRec ignores the refine interaction order information of correlated user sequences. And Beauty and Games have stronger sequential properties than CDs, resulting in significant improvement in the performance of DGSR over HyperRec on Beauty and Games.
	\item SASRec, HGN, SR-GNN, and TiSASRec achieve better performance than neural methods GRU4Rec+ and Caser. One possible reason is that they could explicitly capture the item-item relations by employing attention or hierarchical gating mechanism. Caser generally achieves better performance than GRU4Rec+ in most cases. Such improvement might be attributed to the CNN module, which could capture the more complex behavior pattern than the GRU net. Compared with the excellent performance in the session-based recommendation scenario, the performance of SR-GNN is flat in the sequential recommendation. One possible reason that the lack of repetitiveness of our data, making it challenging for the user sequence to form a graph structure. 
	
	\item BPR-MF achieves poor performance on three datasets. Since BPR-MF can only capture users' general interests, it is challenging to model the user's behavior sequence. GRU4Rec+ slightly underperforms FPMC in Beauty and Games while performing better in CDs. The reason might be that FPMC focuses on dynamic transitions of items, so they perform better on sparse datasets \cite{li2020time}. 
\end{itemize}

\subsection{Study of Dynamic Graph Recommendation Networks (RQ2)}
To investigate DGRN component's superiority in DGSR, we compare DGSR with different variants on Games, Beauty and CDs datasets, which set the various modules for encoding long-term and short-term information. We show the variant models and their results in Table \ref{ablation study} and have the following findings:

\begin{itemize}
	
	\item DGSR-D outperforms DGSR-R and DGSR-G in Games and CDs datasets. We attribute the improvement to the combination of attention mechanism and relative-order embedding, which could adequately distill the long-term information from neighbors of each node. DGSR-R achieves competitive results in Beauty. The reason might be that the length of sequence is small, GRU net could model their dependencies of sequence like dynamic attention module. The GCN-based variants achieves poor performance on three datasets. It is probably because the GCN module treats all neighbor nodes as equally important, which introduces more noise in message propagation. DGSR-A also performs better than DGSR-L. It verifies that only utilize the last interaction embedding is insufficient to capture the short-term information. 
	
	
	\item All variants with two modules (long-term and short-term) are consistently superior to variants with single module (long-term or short term). It illustrates the necessity of combining long-term and short-term information. Although DGSR-R performs better than DGSR-D in Beauty, DGSR-DA is superior to DGSR-RA and DGSR-RL. One possible reason is that DGSR-DA considers the relationship between central node and neighbor nodes, which is conducive to information propagation in the dynamic graph. In contrast, DGSR-RL and DGSR-RA only focus on the interactions of neighbors and ignore the roles of central node.

	
\end{itemize}

\subsection{The Sensitivity of Hyper-parameters (RQ3)}
To explore the effect of explicit modeling dynamic collaborative information among user sequences on DGSR, we study how three hyperparameters, the DGRN layer number $l$, the order of sub-graph, and the maximum length of user sequence $n$ , affect the performance of DGSR. 
\subsubsection{Effect of DGRN Layer numbers} 
\begin{figure}[t]
	\centering
	\subfloat[Games]{\includegraphics[scale=0.5]{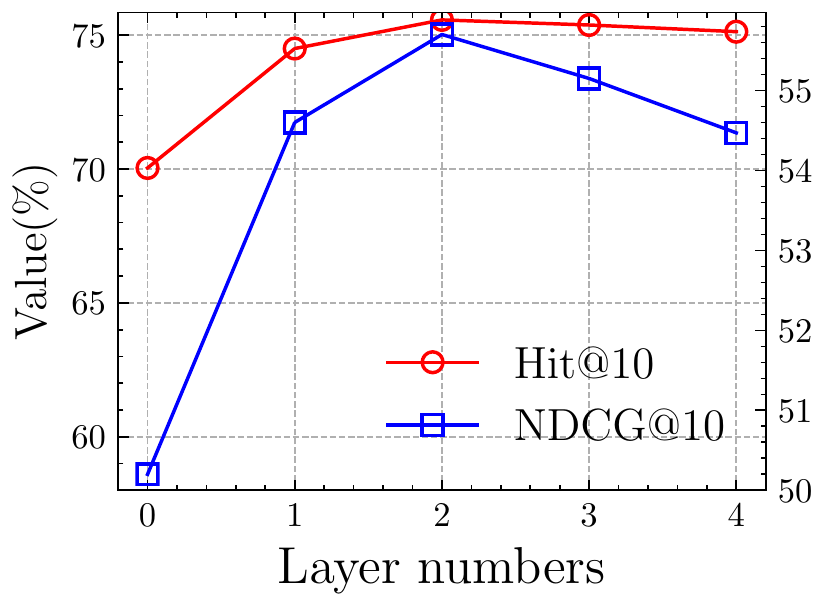}
	}
	\subfloat[Beauty]{\includegraphics[scale=0.5]{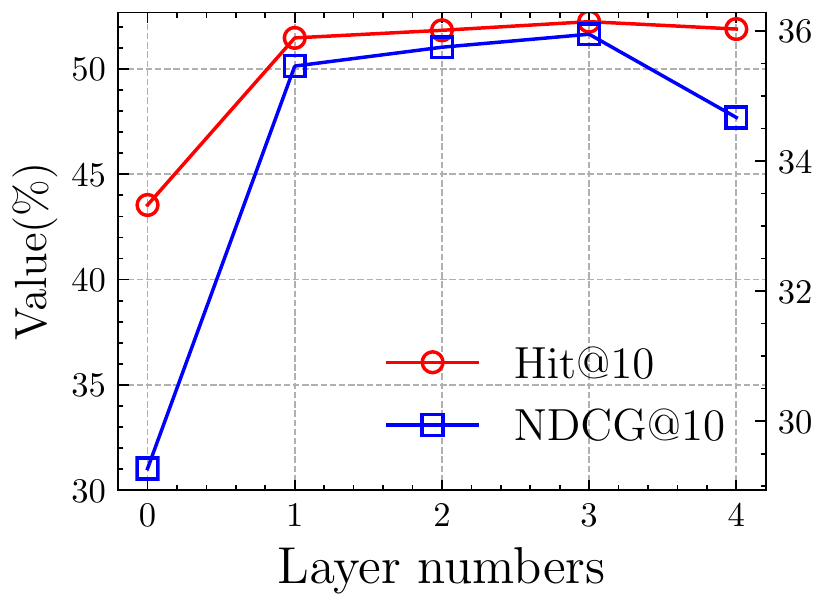}
	}
	\caption{Effect of propagation layer numbers (the y-axis on the left is Hit@10 value, and the right is NGCD@10 value)
	}
	\label{layer}
\end{figure}

We conduct our method with different DGRN layer number $l$ on Games and Beauty data set. DGSR-0 represents only use user embedding and last item embedding for recommendation. DGSR-1 represents the DGRN with one layer, indicating to use only intra sequence information for prediction. DGSR-$l$ ($l$$>$1) indicates DGSR could utilize $l$-order user sequence information to make predication. From Figure \ref{layer}, we find that: 
\begin{itemize}
	\item Increasing the layer of DGSR is capable of promoting the performance substantially. It demonstrates that exploiting high-order user sequences information explicitly can effectively improve recommendation performance. DGSR-2 and DGSR-3 achieve the best performance on Games and Beauty, respectively. One possible reason is that Beauty is sparser than Games, a larger number of layers may be required to introduce a more contextual information.
	
	\item When further stacking propagation layer, we find that the performance of DGSR-3 and DGSR-4 begin to deteriorate. The reason might be that the use of far more propagation layers may lead to over smoothing, which is also consistent to the findings in \cite{li2018deeper}.
	
	\item DGSR-1 consistently outperform DGSR-0 in all cases, even outperforms most baselines. We attribute to the power of the message propagation mechanism in dynamic graph recommendation networks, which could effectively encode the order information in user sequences to extra users' dynamic preferences accurately.
\end{itemize}

\subsubsection{Effect of the sub-graph sampling size}
\begin{figure}[t]
	\centering
	\subfloat[Games]{\includegraphics[scale=0.5]{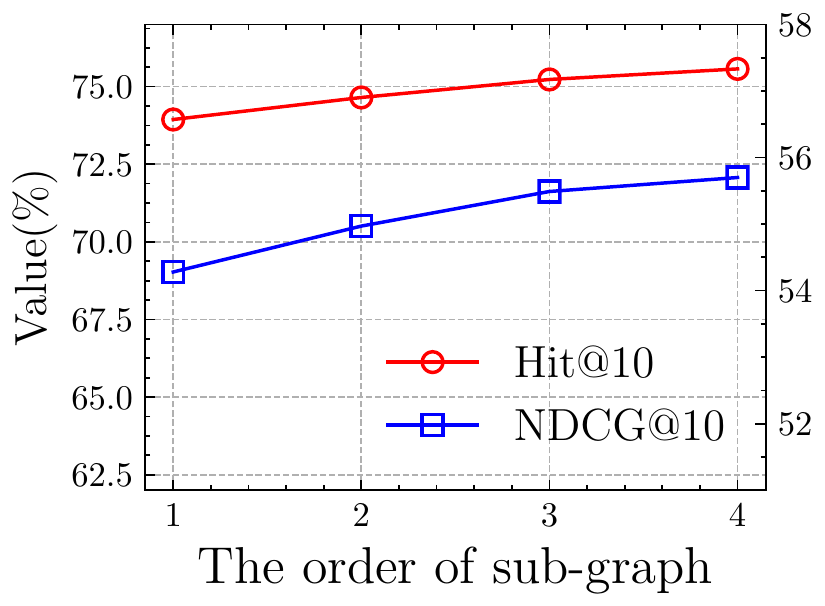}
	}
	\subfloat[Beauty]{\includegraphics[scale=0.5]{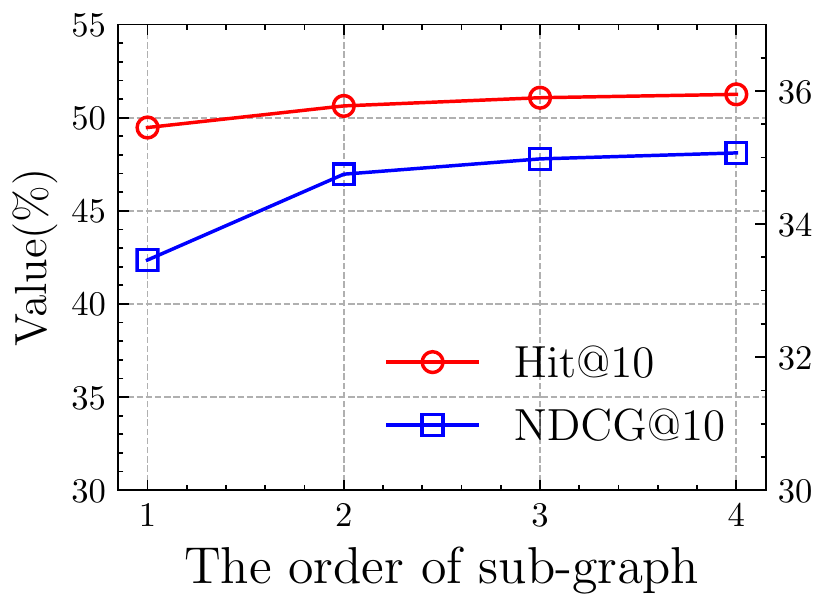}
	}
	\caption{Effect of the sub-graph sampling size (the y-axis on the left is Hit@10 value, and the right is NGCD@10 value)}
	\label{sampling}
\end{figure}

We conduct our method with different sub-graph sampling size. The order of sub-graph $m$ determines the size of the sampling. In particular, we search the $m$ in the range of $\{1,2,3,4\}$. The results in Figure \ref{sampling} show that when $m$ is increased from 1 to 3, the model performance can be effectively improved. The reason is that a larger-sized sub-graph can provide more dynamic contextual information for each user sequence to assist in prediction. With the increase of $m$, the model performance tends to be stable because of the limitation of the number of DGRN layers. In practice, we cannot blindly increase the value of $m$ because the sub-graph size increases exponentially with the increase of $m$, which will cause stuck for our training or testing. 

\subsubsection{Effect of the maximum sequence length}
\begin{figure}[t]
	\centering
	\subfloat[Games]{\includegraphics[scale=0.5]{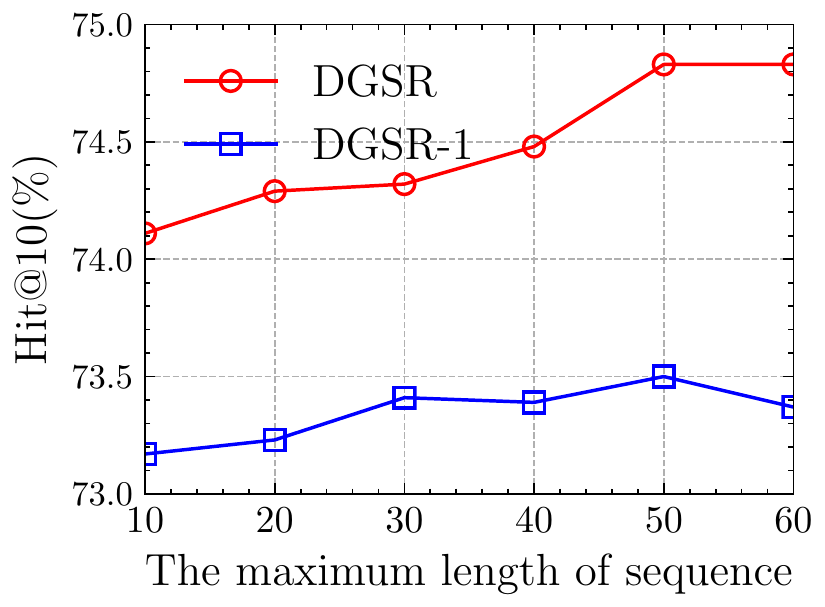}
	}
	\subfloat[Beauty]{\includegraphics[scale=0.5]{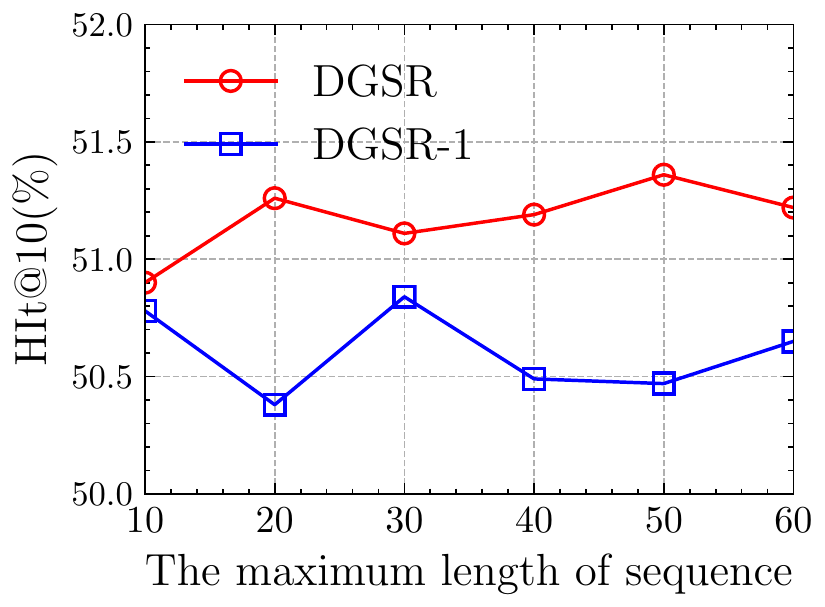}
	}
	\caption{Effect of the maximum length of user sequence}
	\label{length}
\end{figure}

We train and test our method on the Games and Beauty datasets with $n$ from 10 to 60, while keeping other optimal hyperparameters unchanged. Besides, to further investigate the benefit of explicitly utilizing the dynamic collaborative information, we also conduct DGSR-1 with different $n$. Figure \ref{length} shows the Hit@10 results. We have the following findings:
\begin{itemize}
	\item Increasing the $n$ of DGSR from 10 to 50 consistently improves the performance of Games data. DGSR performs better on the beauty when set n to be 20 and 50. However, blindly increasing the $n$ does not necessarily improve the performance of DGSR and DGSR-1. It is likely to bring noise and cause the performance to attenuate. 
	
	\item Compared with DGSR-1, DGSR performs better than DGSR-1 at each value of $n$. To be specific, even when $n$ is set to $10$, DGSR is still better than the best performance of DGSR-1, which implies that explicitly utilizing high-order contextual information of user sequence can alleviate the issue of insufficient user history information, thus improving the performance of recommendation.
\end{itemize}

\subsubsection{Effect of the embedding size}
\begin{figure}[t]
	\centering
	\subfloat[Games]{\includegraphics[scale=0.5]{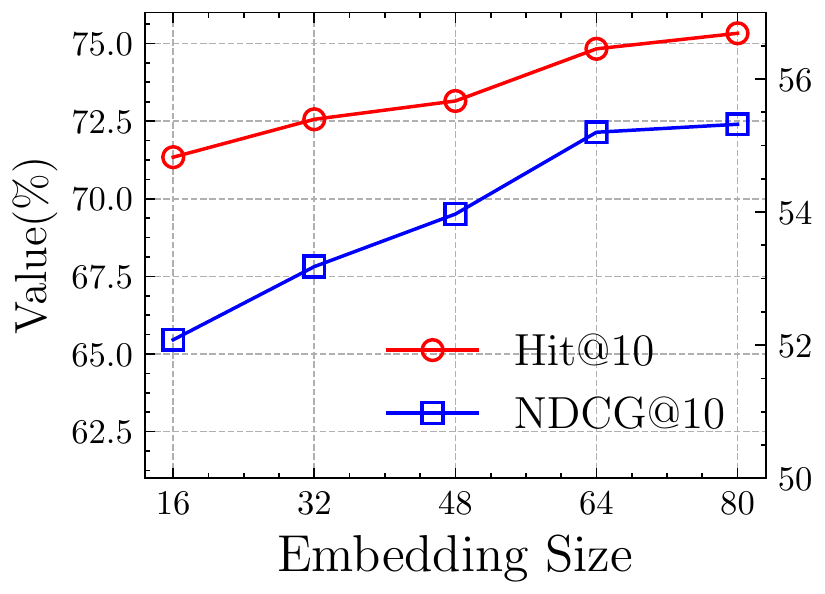}
	}
	\subfloat[Beauty]{\includegraphics[scale=0.5]{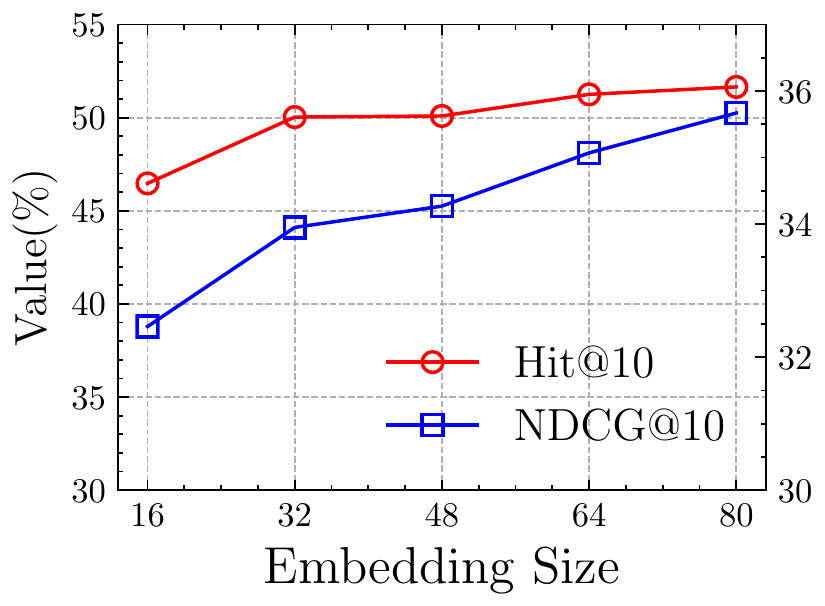}
	}
	\caption{Effect of the embedding size (the y-axis on the left is Hit@10 value, and the right is NGCD@10 value)}
	\label{embedding_size}
\end{figure}

We further analyse the impact of different dimensionality of embeddings. Figure \ref{embedding_size} describes the performance of model under the embedding size from 16 to 80. We can observe that the performance of model gradually improves as the dimensionality increases. With the further increase of the dimensionality, the performance tends to be stable. This verifies the stability of our model in different dimensions.

\section{Conclusion}
This work explores explicitly modeling dynamic collaborative information among different user sequences in sequential recommendation. Inspired by dynamic graph neural networks, we propose a novel method, DGSR. In DGSR, all user sequences are converted into a dynamic graph, which contains the chronological order and timestamps of user-item interactions. The key of DGSR is the well-designed Dynamic Graph Recommendation Network, which realizes the explicit encoding of the dynamic collaborative information among different user sequences. The next-item prediction task is finally converted into a node-link prediction of the dynamic graph so that the model can be trained end-to-end. Extensive experiments on three real-world datasets verify the effectiveness and rationality of DGSR. 

\ifCLASSOPTIONcaptionsoff
  \newpage
\fi



%
\bibliographystyle{IEEEtran}
\bibliography{reference}

\end{document}